%% file: 0.main.tex
\documentclass[manuscript, screen, acmsmall]{acmart}

\setcopyright{none}

\usepackage{eso-pic}
\AddToShipoutPictureBG*{%
  \AtPageUpperLeft{%
    \put(\LenToUnit{0.5\paperwidth},\LenToUnit{-0.93\paperheight}){%
      \makebox[0pt]{{\color{red}\textbf{This is a preprint. Accepted by ACM CSCW 2026}}}%
    }%
  }%
}

\acmDOI{}
\acmISBN{}
\copyrightyear{2026}
\acmYear{2026}
\acmMonth{4}

\acmConference[CSCW '26]{The 29th ACM Conference on Computer-Supported Cooperative Work and Social Computing}{April 2026}{Salt Lake City, Utah, USA}

\usepackage{multirow}
\usepackage{subcaption} 
\usepackage{caption}
\usepackage{array,ragged2e}
\usepackage[inkscapelatex=false]{svg}
\newcolumntype{P}[1]{>{\RaggedRight\arraybackslash}p{#1}}
\definecolor{accessibleblue}{RGB}{30,144,255}
\usepackage{longtable}

\usepackage{setspace}

\usepackage{xcolor}     
\usepackage{letltxmacro} 
\usepackage{xstring}     
\LetLtxMacro{\OriginalTextColor}{\textcolor}
\renewcommand{\textcolor}[2]{#2}

\begin{document}

\title[Analyzing Social Media Claims]{Analyzing Social Media Claims regarding Youth Online Safety Features to Identify Problem Areas and Communication Gaps}

\author{Renkai Ma}
\authornote{Both authors contributed equally to this research.}
\email{renkai.ma@uc.edu}
\affiliation{
    \institution{University of Cincinnati}
  \city{Cincinnati}
  \country{USA}
}

\author{Dominique Geissler}
\authornotemark[1]
\email{d.geissler@lmu.de}
\affiliation{%
  \institution{LMU Munich \& Munich Center for Machine Learning}
  \city{Munich}
  \country{Germany}
}

\author{Stefan Feuerriegel}
\email{feuerriegel@lmu.de}
\affiliation{%
  \institution{LMU Munich \& Munich Center for Machine Learning}
  \city{Munich}
  \country{Germany}
}

\author{Tobias Lauinger}
\email{lauinger@nyu.edu}
\affiliation{%
  \institution{New York University}
  \city{NY}
  \country{USA}}

\author{Damon McCoy}
\email{mccoy@nyu.edu}
\affiliation{%
  \institution{New York University}
  \city{NY}
  \country{USA}}

\author{Pamela Wisniewski}
\email{pwisniewski@icsi.berkeley.edu}
\affiliation{%
  \institution{International Computer Science Institute}
  \city{Berkeley}
  \country{USA}
}

\begin{abstract}
Social media platforms have faced increasing scrutiny over whether and how they protect youth online. While online risks to children have been well-documented by prior research, how social media platforms communicate about these risks and their efforts to improve youth safety have not been holistically examined. To fill this gap, we analyzed $N=352$ press releases and safety-related blogs published between 2019 and 2024 by four platforms popular among youth: YouTube, TikTok, Meta (Facebook and Instagram), and Snapchat. Leveraging both inductive and deductive qualitative approaches, we developed a comprehensive framework of seven \textit{problem areas} where risks arise, and a taxonomy of safety features that social media platforms claim address these risks. Our analysis revealed uneven emphasis across problem areas, with most communications focused on Content Exposure and Interpersonal Communication, whereas less emphasis was placed on Content Creation, Data Access, and Platform Access. Additionally, we identified three problematic communication practices related to their described safety features, including discrepancies between feature implementation and availability, unclear or inconsistent explanations of safety feature operation, and a lack of evidence regarding the effectiveness of safety features in mitigating risks once implemented. Based on these findings, we discuss the communication gaps between risks and the described safety features, as well as the tensions in achieving transparency in platform communication. Our analysis of platform communication informs guidelines for responsibly communicating about youth safety features.

\end{abstract}



\received{20 February 2007}
\received[revised]{12 March 2009}
\received[accepted]{5 June 2009}

\maketitle

\input{1.Intro}

\input{2.RelatedWork}
\input{3.Methods}

\input{4.Findings}

\input{5.Discussion}

\input{6.LimitFuture}
\input{7.Conclusion}

\begin{acks}
We thank the senior program committee (PC) member, the PC member, and all reviewers for their constructive feedback. This work was supported in part by the German Research Foundation (Grant: 543018872), joint internal funding from NYU and Ludwig Maximilian University of Munich, and the Democracy Fund. Additionally, Dr. Wisniewski’s research on youth online safety is supported by the U.S.\ National Science Foundation under grants \#IIS-2550812, \#TI-2550746, and \#CNS-2550834, and by the William T.\ Grant Foundation grant \#187941. Any opinions, findings, and conclusions or recommendations expressed in this material are those of the authors and do not necessarily reflect the views of the research sponsors. We disclose that Drs.\ Damon McCoy and Pamela Wisniewski serve as expert witnesses for the plaintiffs in lawsuits against some of the social media platforms included in this study.
\end{acks}

\bibliographystyle{ACM-Reference-Format}
\bibliography{references}

\newpage
\appendix

\section{Codebook for RQ2: Assessing Problematic Communication Practices}
\label{reflexive_codebook} 
\begin{tiny}
\begin{longtable}{
    >{\raggedright\arraybackslash}p{0.22\linewidth} 
    >{\raggedright\arraybackslash}p{0.23\linewidth} 
    >{\raggedright\arraybackslash}p{0.5\linewidth}  
  }

\caption{Codebook for reflexive thematic analysis of the implementation or effectiveness of their described youth safety features} \\
\\

\toprule
\textbf{Theme} & \textbf{Sub-theme} & \textbf{Example Quote(s)} \\
\midrule
\endfirsthead 

\toprule
\multicolumn{3}{c}%
{{\bfseries \tablename\ \thetable{} -- continued from previous page}} \\
\midrule
\textbf{Theme} & \textbf{Sub-theme} & \textbf{Example Quote(s)} \\
\midrule
\endhead

\midrule
\multicolumn{3}{r@{}}{\textit{Continued on next page}} \\
\endfoot 

\bottomrule
\endlastfoot

\multirow{3}{=}{\parbox{\linewidth}{\textbf{Safety Feature Implementation Claims Often Mismatch Rollout Reality}}} 
  & Provisional and evolving language makes it unclear whether features are implemented
  & \textit{(2024 Snap):} "Parental tools: Soon, in Family Center... parents will be able to see if their 16- and 17-year old teens have an active public story..." \\
  & 
  & \textit{(2024 Snap):} "...Families will soon be able to choose up to three specific locations... and parents will receive notifications when their family member departs from or arrives..." \\
\cmidrule{2-3} 
  & Stated geographic limits make it unclear where features are implemented
  & \textit{(2024 Meta):} "Alternate Topic Nudge Teens under 18 in certain countries will see a notification that encourages them to switch to a different topic if they’ve been looking at the same type of content on Explore for a while." \\

\midrule

\multirow{12}{=}{\parbox{\linewidth}{\textbf{How Platforms' Safety Features Operate Remains Unclear and Inconsistent}}} 
  & The conditions triggering safety features to activate remain unclear and inconsistent
  & \textit{(2023 Snap):} "If you’re a parent over the age of 25, you can use Family Center..." \\
  & 
  & \textit{(2024 Snap):} "Family Center will provide parents over the age of 25 the ability to: See which Snapchat friends..." \\
\cmidrule{2-3}
  & How safety features intersect with each other remains unclear
  & \textit{(2023 Snap):} "...Easily and confidentially report any concerns directly to our Trust and Safety team" (Context is on Family Center). \\
  & 
  & \textit{(2024 Snap):} "When you report an account through Family Center, either from the ‘Recent Conversations’ section or from the ‘Friends List,’ you are taken to the account reporting page..." \\
  & 
  & \textit{(2024 Snap):} "...Easily and confidentially report any accounts parents may be concerned about directly to our 24/7 Trust and Safety team." \\
\cmidrule{2-3}
  & Rationales for varying restriction levels regarding similar features remain unclear.
  & \textit{(TT):} "...only people over 18 are allowed to host a livestream." \\
  & 
  & \textit{(TT):} "Creators must be 18+ to go LIVE..." \\
  & 
  & \textit{(TT):} "your teen has the control to decide who can make duets with or react to their content." \\
  & 
  & \textit{(TT):} "we give you the option to decide who can make duets with or react to your content." \\
\midrule

\multirow{5}{=}{\parbox{\linewidth}{\textbf{How Platforms Present Safety Features Fails to Exhibit Effective Risk Mitigation}}} 
  & How platforms communicate about risks leaves the targeted risks unclear
  & \textit{(Meta):} "...The Sensitive Content Control has three options... “Standard” is the default state... “More” enables people to see more sensitive content.." \\
  & 
  & \textit{(Meta):} "...you may see content that doesn’t break the rules, but could be upsetting... you can adjust the Sensitive Content Control..." \\
\cmidrule{2-3}
  & Safety features’ effectiveness in addressing risks lacks evidence
  & \textit{(YT, 2019):} "suspend comments on videos featuring young minors... we had been working on an even more effective classifier... We accelerated its launch and now have a new comments classifier... that will detect and remove 2X more individual predatory comments." \\
\cmidrule{2-3}
  & How platforms claim safety features work mismatches the scope of the risks they claim to address
& \textit{(YT, 2020):} "As a YouTube creator, you are required to set future and existing videos as made for kids or not... This will help ensure that we offer the appropriate features on your content." \\
  & 
  & \textit{(YT, 2020):} "Regardless of your location, you’re legally required to comply with the Children’s Online Privacy Protection Act... You’re required to tell us that your videos are made for kids..." \\

\end{longtable}
\end{tiny}

\end{document}

%% file: 1.Intro.tex
\section{INTRODUCTION}

In recent years, protecting youth\footnote{In this paper, we use “youth” to refer to children in different age groups, including under 13, teens, and tweens, and we use the terms children and youth interchangeably.} online has become a major focus of regulatory efforts in the United States (U.S.) \cite{FederalTradeCommission1999Childrens312, InformationCommissionersOffice2021AgeServices, FederalTradeCommission2023FTCData, FederalTradeCommission2019GoogleLaw} and platform governance \cite{Xiao2022SensemakingHarm, Razi2023SlidingYouth, Ma2024LabelingYouTube}. For example, in 2019, the U.S. Federal Trade Commission (FTC) fined YouTube and Musical.ly (now TikTok) for violating the Children’s Online Privacy Protection Act (COPPA), which prohibits collecting personal information from children under 13 without parental consent \cite{FederalTradeCommission2019GoogleLaw, FederalTradeCommission2019Musical.lyInc.}. Since then, U.S. lawmakers have introduced new bills, such as the Kids Online Safety Act (KOSA), that seek to hold platforms accountable for protecting minors from online risks such as cyberbullying, data misuse, and the negative impacts of algorithmic content recommendations on social media \cite{Senate2023S.1409Act}. 

Indeed, protecting youth online has been an ongoing research effort for over a decade. The foundational 4Cs risk framework, developed by Livingstone et al. \cite{Livingstone2021TheChildren}, has shaped the field of youth online safety (e.g., \cite{Shin2017HowTeachers, Rodriguez-de-Dios2018AOpportunities}) by categorizing online risks to children into four over-arching categories: \textit{content} (exposure to harmful or inappropriate material), \textit{contact} (unwanted or risky interactions with adults), \textit{conduct} (youths' own risky behavior online), and \textit{contract} (risks of data and commercial practices). The 4Cs framework, which has been updated over the years, has provided a shared taxonomy for understanding the risks that youth face online and has been widely adopted by HCI and CSCW researchers to understand online risks and (re)design safety solutions for youth (e.g., \cite{Badillo-Urquiola2024TowardsCare, Agha2023StrikePrevention}), such as evaluating parental control apps \cite{Wang2021ProtectionSafety} and co-designing new safety nudges with teens to address privacy breaches on social media \cite{Obajemu2024TowardsNudges}.

Besides these \textit{external} safety mechanisms, HCI and CSCW researchers have recently begun examining safety features built \textit{internally} into social media platforms (i.e., built-in features). Much of this emerging work investigates user experiences with specific built-in features on individual platforms, such as parental content filters \cite{Dumaru2024ItsChildren, Dumaru2024ItsTools} and privacy measures \cite{Kumar2017NoPrivate, Ma2024LabelingYouTube} or envisioning new built-in safety features \cite{Agha2023Co-DesigningInterventions, Agha2023StrikePrevention}. To the best of our knowledge, our work is the first to investigate social media platforms' communication about built-in youth safety measures across different types of features and across multiple platforms. While the 4Cs framework \cite{Livingstone2014InOnline, Livingstone2021TheChildren} categorizes online risks that youth encounter, an analogous taxonomy for categorizing safety measures does not currently exist, which is a notable gap given that children use multiple social media platforms every day \cite{Anderson2023TeensSocial2023, CommonSenseMedia2022The2021}. Therefore, understanding the contexts where the known online risks arise to youth and where platforms claim they apply safety features requires a better structure to supplement the current risk-centric 4Cs framework, and we conceptualize these contexts as \textit{problem areas}. 

Industries (e.g., tobacco industry \cite{OConnor2017PerceptionsPurchase, Pepper2014EffectsCues, Ashley2012TheScience}) confronted with public concerns of risks in their products tend to perform various communication practices to minimize the risks and thus not lose customers or lawsuits. Similarly, social media platforms often describe and market their safety features through public communication channels like press releases and help blogs (e.g., \cite{TikTok2019TikTokUsers, YouTube2015YouTubeKids, Instagram2024TeenParents}) to address online risks for youth. Given that end-users (youth, parents, or educators) rely on these communications, essentially as corporate communication \cite{Benedetto1999IdentifyingLaunch, Hultink1997IndustrialPerformance}, to understand and utilize safety features, their clarity, consistency, and actionability are important. From this standpoint, we aim to evaluate the extent to which platform communications especially \textit{misalign} with well-documented online risks (such as those categorized by the 4Cs framework) or the described reality of their features, including feature implementation and effectiveness. When these communications misalign with risks or present inconsistent information, they might hinder user understanding, create a false sense of security among users, including youth, parents, policymakers, or researchers, and impede the effective adoption of safety features in certain problem areas. Thus, we ask two questions to characterize problem areas and uncover issues in communication practices:

\begin{itemize}
    \item \textbf{RQ1:} \emph{What problem areas characterize the contexts where the known risks arise to youth on social media platforms, and where platforms claim to apply safety features?}
    \item \textbf{RQ2:} \emph{What are the problematic communication practices that exist within platform communications regarding youth safety features?}
\end{itemize}

To answer these two research questions, we first selected the four most popular social media platforms among youth in the U.S.: Meta (Instagram and Facebook\footnote{While we recognize that Facebook and Instagram might have different features and user bases, for the purposes of this initial selection, we grouped them under the parent company, Meta, due to shared ownership. Also, in later data analysis, we found that there were not many articles specifically discussing features exclusive to Facebook, as most situations are relevant to both platforms, further supporting this grouping.}), YouTube, Snapchat, and TikTok, and then adopted both inductive and deductive qualitative approaches to collect and analyze $N=352$ press releases and safety-related articles published between 2019 and 2024, which are public channels through which social media platforms formally communicate about their safety features with users.\footnote{The dataset can be found in a GitHub repository: \url{https://anonymous.4open.science/r/youth_online_safety_communication-2805/README.md}}. We inductively coded the articles and then deductively mapped out the described safety features to the known risks to develop a taxonomy of safety features around the identified problem areas. Here, we extended the established 4Cs risk framework to 5Cs and added circulation risk as a new type of risk to cover emerging cases where children's content is inappropriately amplified or propagated on social media platforms. 

We uncovered an uneven distribution of platform communications across problem areas, where most of the communications focus on the Content Exposure and Interpersonal Communication, while focusing less on Platform Access, Monetization, and Content Creation areas (RQ1). Through reflexive thematic analysis \cite{Braun2019ReflectingAnalysis} on our collected articles, we identified three problematic practices within a recognizable lifecycle of platforms’ communication about youth online safety, including (1)~discrepancies between platform implementation claims and feature availability, (2)~lack of clarity and consistency in explaining feature operation, and (3)~unclarity and lack of evidence in demonstrating effectiveness of risk mitigation (RQ2). These findings motivate us to discuss the gaps between risks and described safety features and the tensions in achieving transparency in platform communication practices. We further discuss the communication implications of the alignment and contrast between the design philosophies suggested by current platform communication practices and those supported by prior research on youth online safety and regulatory considerations. We then lay out guidelines for responsible platform communications regarding youth safety features. 

Our paper makes four primary contributions. First, we extend the established 4Cs risks framework to 5Cs to cover emerging cases of \textit{circulation risk}. Second, we introduce the concept and framework of \textit{problem areas} to characterize contexts where youth might encounter known risks and where platforms claim to apply safety features to address those risks. This framework supplements the current understanding of online risks by addressing the relative lack of work in taxonomizing and categorizing online safety features for youth, especially those built internally into social media platforms. Third, to the best of our knowledge, our study is the first to both identify what safety features the most popular social media platforms claim to implement, and to examine the communication practices by social media platforms around those features as part of their youth online safety efforts. By identifying problematic practices in this communication, our work aligns with the current trend in HCI and CSCW toward evaluating platforms' safety features (e.g., \cite{Obajemu2024TowardsNudges, Agha2023StrikePrevention, Badillo-Urquiola2024TowardsCare, Caddle2025BuildingSafety}). Lastly, based on our analysis of platform communication, we lay out both communication implications for youth online safety designs as well as communication guidelines for responsibly communicating about safety features.

%% file: 2.RelatedWork.tex
\section{BACKGROUND}
We review how prior work examines online risks to children and safety features that address those risks on social media platforms. Like many industries facing public scrutiny of risks in their products, social media platforms respond to it through different communication practices to retain users and avoid lawsuits. Given the limited work assessing these practices, our study is the first to fill this gap. Finally, we introduce the overview of safety products on social media platforms. 

\subsection{Youth, Social Media Use, and Online Risks}
Social media platforms like YouTube, TikTok, Instagram, and Snapchat have seen a notable rise in usage among youth in the U.S., according to Pew Research Center \cite{Anderson2023TeensSocial2023, Faverio2024Teens2024}. Several national surveys also indicate that using platforms such as YouTube or TikTok has been the most popular media activity among eight- to eighteen-year-olds \cite{Moyer2022KidsFinds, CommonSenseMedia2022The2021}. These platforms offer various benefits for children: YouTube supports children’s informal learning \cite{Imaniah2020YoutubeBehaviors} and self-expression through content creation and sharing \cite{McRoberts2016DoYouTubers, Ruiz-Gomez2022PlayingInfluencers}. Also, TikTok immerses youth in contemporary youth culture and helps them form online communities \cite{Zulli2022ExtendingPlatform}.

However, youth also encounter a variety of risks on social media platforms. Teens might face harmful content (e.g., violent content \cite{Livingstone2014InOnline}) and cyberbullying \cite{Ashktorab2016DesigningTeenagers, Schoenebeck2021YouthJustice}, or become victims of privacy breaches \cite{Masaki2020ExploringThreats, Alemany2019EnhancingMechanisms}. Researchers have found that younger children, specifically those under 13, lack technical literacy and perceive data privacy risks in simple terms, such as “stranger danger” \cite{Kumar2017NoPrivate}. They have limited awareness of algorithmic practices like data tracking or targeted ads, and often feel powerless to control data collection by social media platforms \cite{Zhao2019IOnline, Wang2022DontOnline}.

Livingstone et al. have helped categorize these diverse risks into a 4Cs framework, comprising of \textit{content}, \textit{contact}, \textit{conduct}, and \textit{contract} risks~\cite{Livingstone2021TheChildren, Livingstone2014InOnline}. Content risks refer to exposure to harmful material, such as violent, hateful, or sexual content that is age-inappropriate. Contact risks describe harmful adult-initiated interactions, such as grooming, harassment, or sexual exploitation. Conduct risks involve negative peer interactions among children (e.g., bullying, hateful behaviors). Lastly, contract risks concern exploitation through commercial practices (e.g., scams, targeted ads). 

The 4Cs framework has been widely adopted by researchers to categorize online risks to youth on social media platforms (e.g., \cite{Lee2013ParentalWhom, Zhao2022UnderstandingInstagram, Rodriguez-de-Dios2018AOpportunities}). For example, prior work found that contact risks emerge on Instagram, where teens report frequent sexual solicitation and harassment that increases if they ignore the solicitations \cite{Dev2022FromConversations}. Meanwhile, contract risks can emerge from social media platforms' commercial practices. For example, YouTube's opaque, ad-driven recommendation algorithms have been found to exploit younger children through misleading product promotion videos \cite{Boerman2016InformingContent, Evans2018ParentingVideos} and attract them to watch such videos for extended periods \cite{DeVeirman2019WhatResearch, Boerman2020DisclosingRelationship}. To address these different risks, CSCW research, such as \citet{Wang2021ProtectionSafety}, has applied the 4Cs framework to evaluate safety mechanisms like parental control apps, whereas \citet{Agha2023StrikePrevention} and \citet{Obajemu2024TowardsNudges} have co-ideated and designed online safety interventions together with teens. Our study thus aims to build on the 4Cs framework to understand and categorize the known risks to youth on social media.

\subsection{Mapping Online Risks to Safety Measures to Protect Youth on Social Media}
Recent HCI and CSCW work has acknowledged the role of safety measures or features in addressing online risks for youth, while much of this work has focused on (re)designing them. Researchers, for example, have engaged teens in co-designing privacy management tools to address privacy breaches from other users or the social media platforms themselves \cite{Agha2023StrikePrevention, Wang2023TreatOnline, Badillo-Urquiola2021ConductingAdolescents}. Another line of work has explored using in-situ social media data donations from teens to develop machine learning systems that identify various conduct and content risks \cite{Razi2023SlidingYouth, Ali2023GettingYouth}.

Different from these \emph{external} safety measures (e.g., parental control apps typically not developed by social media companies \cite{Wisniewski2017ParentalSafety, Wang2021ProtectionSafety}), researchers have recently increasingly examined user experiences with safety features built \textit{internally} into social media platforms (i.e., \emph{built-in} features). For example, Dumaru et al. explored \cite{Dumaru2024ItsChildren} and redesigned~\cite{Dumaru2024ItsTools} parental controls on YouTube, enabling parents to limit certain content categories (e.g., war, drugs, alcohol), while further uncovering that these controls reduced exposure to inappropriate content and further promoted parent-child discussions. 
On Instagram, teens report dissatisfaction with built-in reporting tools, criticizing their ineffectiveness in addressing harassment and misinformation \cite{Agha2023Co-DesigningInterventions, Agha2023StrikePrevention}. Similarly, Edelson experimented with algorithmic feed recommendations on Instagram Reels and TikTok for both 13- and 23-year-old female personas, finding similarly sexually suggestive content on Reels for both personas, while TikTok showed less explicit content to the teen than the adult \cite{Edelson2024TeenMedia}.

Despite the user experience with individual built-in safety features, a step back is needed to holistically understand what safety features are available, as claimed by the platforms. While the 4Cs framework \cite{Livingstone2014InOnline, Livingstone2021TheChildren} helps categorize online risks, an analogous taxonomy of safety measures for assessing how platforms claim, or tend, to address youth online safety does not currently exist, particularly given that children use multiple platforms daily \cite{Anderson2023TeensSocial2023, CommonSenseMedia2022The2021}. Our study thus addresses this gap by providing a taxonomy of safety features along the \textit{problem areas} to describe contexts, where risks arise, and where platforms claim to implement the features to address those risks.

\subsection{Communication Practices of Social Media Platforms for Youth Online Safety}
Industries confronted with rising public awareness of risks in their products tend to perform various communication practices to minimize or deflect these risks and thus not lose customers or lawsuits. For instance, the tobacco and e-cigarette industries have historically used terms like ``light'' or ``safer alternatives'' to downplay health risks, fostering consumer perceptions of reduced harm and increasing product appeal \cite{OConnor2017PerceptionsPurchase, Pepper2014EffectsCues, Ashley2012TheScience}. Similarly, communication strategies emphasizing healthfulness or utility for quitting cigarettes correlate with higher youth adoption of e-cigarettes \cite{Pokhrel2015ReceptivityUse, Collins2019E-CigaretteInformation}. These parallels reflect how companies try to use their communication to align business goals with public well-being \cite{Porter2006StrategyResponsibility}. However, such communication often triggers skepticism \cite{Rim2016DimensionsCSR} when companies' one-way communication prioritizes publicity over dialogue \cite{Grunig1995ModelsSetting}.

As related work on online risks to children has highlighted, social media platforms now also face challenges regarding the perceived safety of their platforms. These platforms have implemented (what they claim are) safety measures, which they market or communicate to users through press releases and help blogs (e.g., \cite{TikTok2019TikTokUsers, YouTube2015YouTubeKids, Instagram2024TeenParents}). They begin marketing features directly to users, such as Instagram promoting its Teen Accounts via TV and podcasts \cite{Barwick2025AsOffensive}, and TikTok promoting its reporting tools to prevent harmful behavior during international football games \cite{TikTok2024TikTokNations}. However, discrepancies often surface between platform claims about these features and their effectiveness in protecting children. For example, a Kentucky lawsuit alleges TikTok's screen time controls were designed primarily for public relations rather than reducing teens' usage \cite{AssociatedPress2024TikTokShows}. Instagram’s parental supervision options also face criticism for assuming all parents can navigate social media's complexities, which is not always true \cite{Baxter2024NewOnline}.

To the best of our knowledge, our study is the first to examine how platforms communicate about their built-in safety features, particularly around youth online safety. Filling this gap is critical, as platforms' communication strategies shape public perception, user adoption, and, ultimately, youth online safety. Therefore, our analysis focuses on how platforms communicate safety measures across risk categories and includes a critical examination of their communication practices.

\subsection{Historical Overview of Safety Products on Social Media Platforms}
Over the past decade, major social media platforms have launched various safety products to address online risks for youth. YouTube launched YouTube Kids in 2015 as a standalone, curated platform for younger audiences. In 2019, TikTok introduced TikTok for Younger Users, a separate app experience designed for children under 13 with limited functionality and stricter content controls. More recently, both Meta and Snapchat rolled out Family Center tools in 2022, providing parents with dashboards to monitor youth activity and adjust privacy or content settings. 

\begin{figure}[ht]
    \centering
    \includegraphics[width=0.85\linewidth]{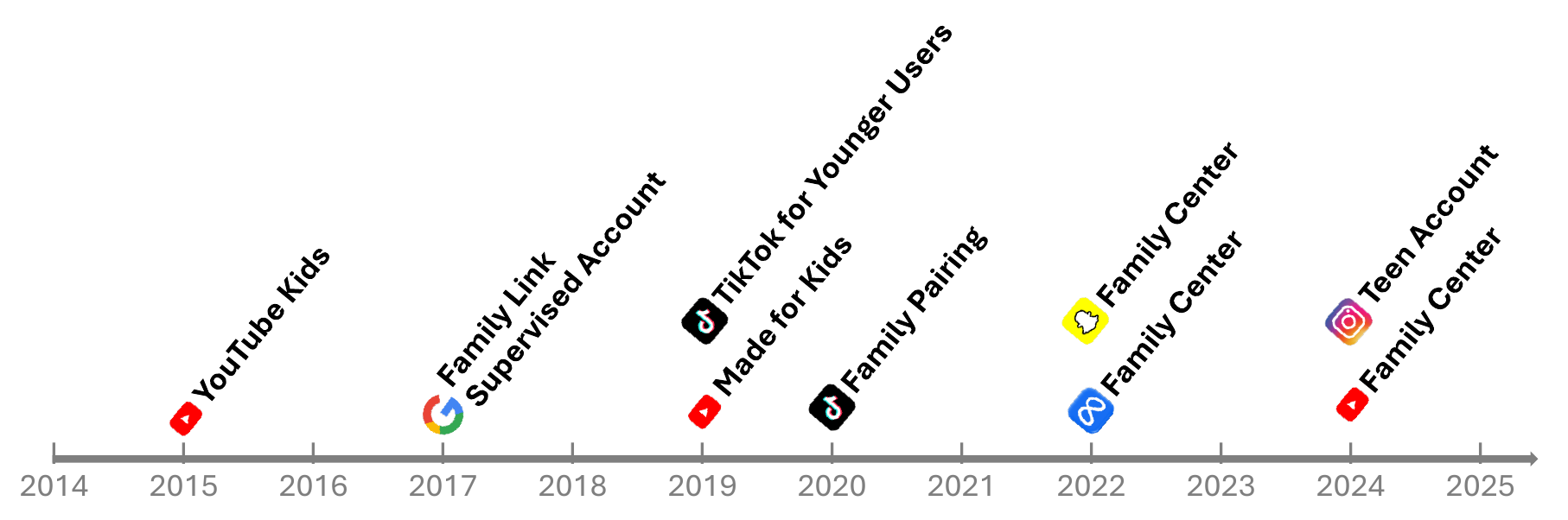}
    \caption{Historical account of overarching youth safety products that social media platforms include in their communications. We included Google here, as YouTube claims to use Google products to safeguard youth online.}
    \label{fig:unique_product_timeline}
\end{figure}

Figure~\ref{fig:unique_product_timeline} provides an overview of these products over time. While these initiatives reflect efforts to respond to public and regulatory pressure, they also serve as strategic communication efforts to position platforms as proactive and responsible. Our findings build on this context by analyzing how such efforts are described at a more granular level, through the safety features they claim to address online risks, and how they communicate about their safety features.

%% file: 3.Methods.tex
\section{METHODS}
This section details the collection and preparation of our dataset about social media platform communications about youth online safety. We further introduce our qualitative coding (RQ1) and reflexive thematic analysis (RQ2) to answer our RQs.

\subsection{Data Collection}
\label{sec_datacollection}

We collected data on how social media platforms communicate about youth online safety in four steps: 

Step~1: We selected four representative platforms: YouTube (including YouTube Kids), Meta (representing Instagram and Facebook), TikTok, and Snapchat. These were chosen because they are the most used platforms by youth across age groups in the U.S. \cite{Statista2023Leading2023, Anderson2023TeensSocial2023}. 

Step~2: \textcolor{black}{We collected data directly from these platforms to identify the safety features each claims to offer, focusing on official press releases (Source 1) and Help or Safety Center blogs (Source 2) from the platforms themselves.} Press releases, sourced from official newsrooms like TikTok Newsroom\footnote{\url{https://newsroom.tiktok.com/en-us}} or Snapchat Newsroom\footnote{\url{https://newsroom.snap.com/}}, regularly announce plans, launches, or updates for built-in safety features. We searched these newsrooms for relevant press releases (see Table~\ref{tab:data-collection}, Source 1). Concurrently, platform Help or Safety Centers, such as Meta's Safety Center\footnote{\url{https://about.meta.com/actions/safety}}, publish blogs about using their safety features. We collected these blogs across the centers using the platforms' search function with keywords: youth OR child OR teen OR minor OR underage OR age (see Table~\ref{tab:data-collection}, Source 2). Both data sources are the platforms' official materials for their communication about safety features to end-users\footnote{While our focus is on the direct-to-user communications found in press releases and help/safety centers, we recognize that platforms also produce regulatory documents like transparency reports or risk assessments for laws such as the Digital Services Act (DSA). We intentionally excluded these documents from our analysis to maintain a clear focus on the materials created for and consumed by the general public, including youth and their caregivers.}. 

To collect these articles and blogs, we used Instant Data Scraper\footnote{\url{https://chromewebstore.google.com/detail/instant-data-scraper/ofaokhiedipichpaobibbnahnkdoiiah}}, a Chrome extension. We focused on articles published from January 1, 2019, to December 31, 2024; this period was chosen because 2019 marked the start of major legal settlements between regulators and social media companies (e.g., the U.S. FTC fined YouTube and TikTok for child privacy law violations in 2019 \cite{FederalTradeCommission2019GoogleLaw, FederalTradeCommission2019Musical.lyInc.}). Data collection ended on December 31, 2024. 
After removing pre-2019 articles, we obtained a dataset of $N = 1,952$ press release articles. While press releases (Source 1) typically included timestamps, many Safety/Help Center blogs (Source 2) did not, so we used the Internet Archive's Wayback Machine\footnote{\url{https://web.archive.org/}} to approximate the publication year of each blog by determining its earliest archived appearance and removing any blogs with an earliest archived appearance before 2019. This yielded $N = 376$ blogs.

\begin{table}[htbp]
\centering
\tiny
\caption{Data sources and article counts for each platform: press releases (Source 1) and Safety/Help Center blogs (Source 2). Please note that Instagram and Facebook belong to Meta overall, but we gathered their materials separately.}
\label{tab:data-collection}
\begin{tabular}{p{0.8cm}
                >{\raggedright\arraybackslash}p{2.9cm}  
                p{0.9cm}
                >{\raggedright\arraybackslash}p{2.9cm}  
                p{1cm}
                p{1.5cm}
                p{1.6cm}}
\toprule
\textbf{Platform} 
& \multicolumn{2}{c}{\textbf{Source 1}} 
& \multicolumn{2}{c}{\textbf{Source 2}} 
& \textbf{Article Count After Relevance Coding} 
& \textbf{Total Count After Fetching Article Hyperlinks} \\
\cmidrule(lr){2-3} \cmidrule(lr){4-5}
& \textbf{Press Release} & \textbf{Article Count} 
& \textbf{Safety/Help Center} & \textbf{Blog/Article Count} 
&  
&  \\
\midrule
YouTube 
& \url{https://blog.youtube/news-and-events/} 
& 379 
& \url{https://support.google.com/youtube#} 
& 204 
& 69 
& 134 \\
\midrule
TikTok 
& \url{https://newsroom.tiktok.com/en-us} 
& 886 
& \url{https://www.tiktok.com/safety/en/} 
& 30 
& 67 
& 74 \\
\midrule
Snapchat 
& \url{https://newsroom.snap.com/} 
& 295 
& \url{https://values.snap.com/safety/safety-center} 
& 33 
& 36 
& 45 \\
\midrule
Meta 
& \url{https://about.instagram.com/blog}, \url{https://about.fb.com/news/} 
& 392 
& \url{https://help.instagram.com/}, \url{https://about.meta.com/actions/safety} 
& 109 
& 99 
& 99 \\
\midrule
\textbf{Total} 
&  
& \textbf{1,952} 
&  
& \textbf{376} 
& \textbf{271} 
& \textbf{352} \\
\bottomrule
\end{tabular}
\end{table}

\begin{figure}[ht]
    \centering
    \includegraphics[width=1\linewidth]{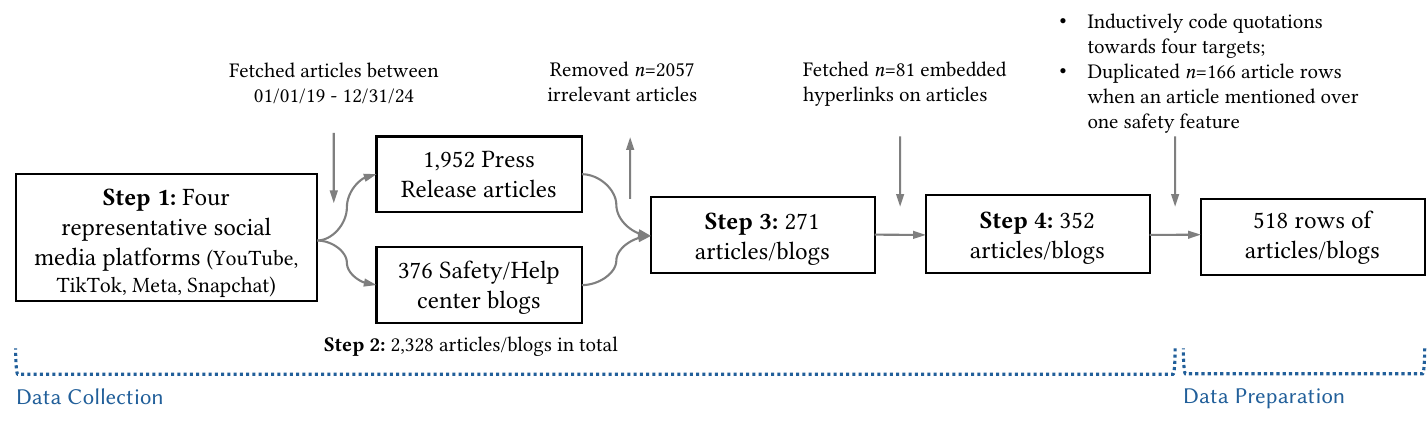}
    \caption{Flowchart of our data collection and preparation.}
    \label{fig:researchprocess}
\end{figure}

Step~3: We performed \textit{relevancy coding} to determine which articles and blogs are relevant to our research questions. We filter our dataset using three inclusion criteria: mentions of (1)~youth (e.g., teens, children, minors), (2)~online safety, and (3)~technical features or implementations. \textcolor{black}{These criteria are applied to each article, and we remove all articles that do not fit the criteria from the dataset.} For example, we excluded articles like TikTok’s “Songs of the Summer 2024”, Instagram’s wildlife exploitation post, and YouTube’s “Get started with channel memberships”, as their topics were unrelated to youth online safety features. To apply these criteria, the dataset was divided between two coders. One coder analyzed the articles from YouTube and Meta, while the second coder analyzed those from TikTok and Snapchat. \textcolor{black}{To validate coding and the consensus of identified relevant articles, we randomly selected five press releases and five Safety/Help Center blogs per platform (\char`\~20\% of data) to be coded by a second reviewer (the other reviewer respectively) and calculated Cohen’s kappa. Inter-rater reliability (IRR) \cite{Cohen1968WeightedCredit} reached substantial (0.61--0.80) or almost perfect (0.81--1) agreement.} For press releases, IRR was: (1)~about youth: 98\% agreement rate, $\kappa=0.953$; (2)~about online safety: 98\%, $\kappa=0.958$; and (3)~mention of technical implementation or features: 88\%, $\kappa=0.689$. For blogs, the values were (1)~96\%, $\kappa=0.9$; (2)~100\%, $\kappa=1$; and (3)~100\%, $\kappa=1$. The two coders resolved disagreements through weekly meetings and joint reviews, producing a final set of $N = 271$ relevant articles (see Table~\ref{tab:data-collection}). 

\textcolor{black}{Step~4: We enriched our dataset through snowball sampling by following embedded hyperlinks in the relevant articles, similar to a backward snowball sampling approach in systematic literature review \cite{Wohlin2014GuidelinesEngineering}. To do so, we went through each relevant article and manually clicked the hyperlinks in it. We included hyperlinks if they are press releases or help and safety center articles of the social media platform. We reapplied the relevancy criteria again and kept those that are about youth, online safety, and a technical feature or implementation. We repeated the snowball sampling up to three levels deep. This way, we collected 81 additional articles for our dataset. This step was done to ensure that we capture as many relevant articles from the social media platforms as possible, because relying solely on our keyword search, we might have missed relevant articles.} Our final dataset contained $N=352$ relevant articles.

\subsection{\textcolor{black}{Data Preparation}}
\label{sec_datanalysis}

\textcolor{black}{After collecting the set of relevant articles, we prepared an analytic dataset that could be used to answer our research questions. To do so, we applied two rounds of qualitative analysis \cite{Braun2006UsingPsychology} to generate codes for our data analysis (both are detailed below). In the first round, we inductively extracted quotations of safety features described in platform communications and coded these into standardized safety features. These form the unit of analysis in our dataset. In this context, “codes” refer to extracted units of meaning, such as safety features and affected age groups of the features. In the second round, we deductively mapped these features into the 5Cs risk framework. These coding rounds produced the feature-level dataset that underpins our subsequent analyses: RQ1 draws on the standardized safety features, while RQ2 draws on the quotations.}

\subsubsection{\textcolor{black}{First-Round Coding of Safety Features (Inductive Approach)}}

\textcolor{black}{The objective of the first round of coding was to inductively extract and structure how platforms communicated safety features in their own articles.} Two coders read each relevant article and identified quotations describing safety features. We added these into our dataset as a \emph{quotation column}. Next, for each quotation, we extracted four items into new dataset columns: (1)~the \textit{described safety feature}, (2)~the \textit{affected age group} by the feature, (3)~the \textit{controlling entity} (i.e., who uses or controls the feature), (4)~any \textit{described sub-feature(s)} (if the primary feature was composite). We oriented our codes to the platform’s original quotation verbatim throughout this inductive coding. For example, from the YouTube quote, “A YouTube Kids profile is a profile within a parent's Google Account. Profiles are available only when the parent is signed in to YouTube Kids with their Google Account,” we coded: (1)~“individual kid profile” as the described safety feature, (2)~“children under 13” as the affected age group, (3)~“parents” as the controlling entity, (4)~no additional sub-features. This inductive round generated an initial dataset of 124 feature codes.

Next, we standardized the wording of the described safety features to establish a uniform categorization of them across the four platforms. For example, Meta’s feature involving technology to detect user age and YouTube’s feature requesting additional proof of age were both consolidated under the standardized feature category “Age Verification.” This standardization process consolidated an initial 124 feature codes into 30 codes across the platforms. This initial data analysis resulted in a dataset of $N=518$ rows, with some articles or blogs repeated due to the feature-level analysis (see the final stage in Figure \ref{fig:researchprocess}).

\subsubsection{\textcolor{black}{Second-Round Mapping to Risk Categories (Deductive Approach)}}

In the second round, the same two coders deductively mapped the described sub-features into the 5Cs risk framework as five new columns in our dataset: \textit{content}, \textit{contact}, \textit{conduct}, \textit{contract}, and \textit{circulation}. We initially used the 4Cs risk framework \cite{Livingstone2008RiskyInternet, Livingstone2008ParentalUse, Livingstone2021TheChildren}. However, as we analyzed the data, particularly considering the rise of youth creating and sharing content on platforms like YouTube and TikTok \cite{McRoberts2016DoYouTubers, Ruiz-Gomez2022PlayingInfluencers, Zulli2022ExtendingPlatform}, we noticed described features restricting the sharing, downloading, or propagation of children’s content (i.e., content by child creators or content featuring children) beyond the intended audience and in potentially harmful ways. This led us to identify an emergent risk category as \textit{circulation} risk and extend the 4Cs framework to 5Cs to capture these cases.

\subsection{Data Analysis}
\subsubsection{Analysis for RQ1: Taxonomizing Safety Features Around Problem Areas}
\label{sec:methods_RQ1}
For RQ1, our goal was to understand the contexts on social media platforms where risks can arise for youth and where platforms communicate that they apply safety measures. To analyze this relationship, we developed a framework of \textit{problem areas}. This was motivated by the need to bridge the gap between existing risk-centric frameworks (like the 4Cs or 5Cs in our study, which categorize what risks exist) and feature-specific work (e.g., experiences with individual built-in safety features \cite{Kumar2017NoPrivate, Dumaru2024ItsChildren, Agha2023StrikePrevention}). 

Our analysis for RQ1 involved two main steps. First, we grouped the 30 described safety features we coded into seven problem areas, which are defined as contexts where risks can arise for youth on social media platforms. Safety features were assigned to a problem area based on which area of user activity they affect. For example, we conceptualized the area of Interpersonal Communication, which covers user activities like messaging and commenting, as a problem area because these activities can expose youth to risks such as online bullying, harassment, or even exploitation. Platforms state that they apply features like Message Restrictions and Connection Controls to mitigate the risks within this area. In addition, we identify a cross-cutting layer of safety features that enable parents to monitor and control their children's activity across the problem areas. We call this layer \textit{Parental Supervision} and group features that involve parents in this additional layer. While we recognize that the list of 30 safety features and the five known risks might not be exhaustive, they were instrumental in helping define the framework of problem areas on social media platforms. 

\begin{small}
\tiny
\begin{spacing}{1}
\begin{longtable}{p{0.25\linewidth}p{0.2\linewidth}p{0.5\linewidth}}
\caption{Codebook for RQ1 used to classify platform safety features and the associated risks by problem area.}
\label{tab:codebook_safety_features}\\
\toprule
\textbf{Problem Area (RQ1)} & \textbf{Described Safety Feature} & \textbf{Feature Definition} \\
\midrule
\endfirsthead

\toprule
\textbf{Problem Area (RQ1)} & \textbf{Described Safety Feature} & \textbf{Feature Definition} \\
\midrule
\endhead

\midrule
\endfoot

\bottomrule
\noalign{\vskip 3pt}
\multicolumn{3}{p{\linewidth}}{\footnotesize *Parental Supervision is not a problem area, but it encompasses many safety features that help parents protect their children against online risks, so we include it here for completeness.}\\

\endlastfoot

\multirow{2}{=}{\parbox{\linewidth}{\textbf{Content Exposure:} \newline Viewing, discovering, and being recommended content, including ads }}
& Content Viewing Levels 
& Control the viewing of content based on topics or age-appropriateness, including ads content \\
\cmidrule{2-3}
& Content Removal 
& Remove harmful or otherwise policy-violating content \\
\cmidrule{2-3}
& Content Recommendation Restrictions 
& Limit visibility/recommendation of policy-violating content such as sensitive, mature, or violent content, but not to the extent of removing it \\
\cmidrule{2-3}
& Reporting 
& Report inappropriate content, accounts, messages, or interactions to the platform \\
\cmidrule{2-3}
& Audience Age Settings 
& Set minimum viewer ages or designate content for kids based on their age for content creators \\
\cmidrule{2-3}
& Content Approval 
& Share or approve selected videos or channels for a child, mostly by parents \\
\cmidrule{2-3}
& Search Controls 
& Control child's search access by disabling search bar and managing history, mostly by parents \\
\midrule

\multirow{2}{=}{\parbox{\linewidth}{\textbf{Interpersonal Communication:} \newline The ways in which users engage in direct or indirect social interaction }}
& Message Restrictions 
& Manage messaging permissions: restrict senders, disable DMs, control group chats \\
\cmidrule{2-3}
& Message Safety Filters 
& Detect, filter, and alert inappropriate content in messages \\
\cmidrule{2-3}
& Connection Controls 
& Manage follower requests and restrict suspicious adults from connections \\
\cmidrule{2-3}
& Content Engagement Restrictions 
& Regulate (e.g., warn, delete, block) comments, likes, tags, and mentions on user content \\
\cmidrule{2-3}
& Interaction Settings Nudge 
& Prompt users to review and potentially update user interaction-related settings, such as comments, mentions, follows, and friends \\
\cmidrule{2-3}
& Reporting 
& Report inappropriate content, accounts, messages, or interactions to the platform \\
\midrule

\multirow{2}{=}{\parbox{\linewidth}{\textbf{Platform Usage Patterns:} \newline Concerning the frequency, duration, and nature of platform use }}
& Time Limit Settings 
& Manage screen time by setting usage limits or time reminders \\
\cmidrule{2-3}
& Notification Settings 
& Pause/mute notifications via quiet modes or scheduled/automatic nighttime limits \\
\cmidrule{2-3}
& Autoplay Controls 
& Enable or disable the automatic playing of the next content \\
\cmidrule{2-3}
& Alternate Topic Nudge 
& Prompt users to switch content topics after prolonged consumption \\
\midrule

\multirow{2}{=}{\parbox{\linewidth}{\textbf{Data Access:} \newline Managing personal data access by others or by the platforms}}
& Data Privacy Controls 
& Set and control personal data privacy such as account visibility, content sharing, location, and discoverability from nonconsensual entities \\
\cmidrule{2-3}
& Privacy Settings Nudge 
& Nudge users to review and update their general privacy settings \\
\midrule

\multirow{2}{=}{\parbox{\linewidth}{\textbf{Monetization:} \newline Financial activities on platforms, such as earning or spending money }}
& Monetization Restrictions 
& Limits monetization features like earning, payments, gifting based on user age \\
\cmidrule{2-3}
& Ads Personalization Settings 
& Manage ad preferences and personalized ads \\
\cmidrule{2-3}
& Purchase Restrictions 
& Limits in-app purchases and virtual gift buying based on age or content type \\
\midrule

\multirow{2}{=}{\parbox{\linewidth}{\textbf{Content Creation:} \newline Creating, posting, and sharing public content (e.g., posts, videos, live streams) }}
& Content Collaboration Restrictions 
& Restricts how others reuse or circulate children's content via Remix, Duet, Stitch, downloads \\
\cmidrule{2-3}
& Account Termination 
& Terminate accounts for violating content policies (e.g., circulating inappropriate content of youth) \\
\cmidrule{2-3}
& Content Creation Restrictions 
& Limits access to certain content creation features (e.g., Live stream) by age \\
\midrule

\multirow{2}{=}{\parbox{\linewidth}{
\textbf{Platform Access:} \newline
 Ability to access a platform's content and functions
}}
 & Age Verification & Verifies user age via inputs, documentation, or detection tools \\
\cmidrule{2-3}
 & Platform Access Controls & Allows caregivers to permit or block platform access \\
\midrule

\multirow{2}{=}{\parbox{\linewidth}{\textbf{Parental Supervision:*} \newline Possibility for parents/guardians to oversee and guide a child's platform use, and we group features that mention ``parent'' or ``parental'' in a literal way here.}}
& Parental Monitoring 
& Parents can monitor youth's online activity, including their connections, settings, communication metadata, and time spent \\
\cmidrule{2-3}
& Parental Control Settings 
& Manage overall parental supervision settings, including control approvals \\
\cmidrule{2-3}
& Individual Kid Profile 
& Allows parents to make distinct profiles with personalized settings for each child \\

\end{longtable}
\end{spacing}\end{small}

Second, we analyzed the co-occurrence of described safety features and the associated 5Cs risks. For each of the five risk types, we aggregated all features coded as addressing that risk and treated this count as 100\% to normalize the distribution. We then calculated the percentage of these features falling into each of the problem areas. These distributions were visualized as a heatmap through Python 3.9 (see Figure \ref{fig:heatmap_problemareas_risks_features}), where columns represent the 5Cs risk types, rows represent problem areas, and cell values indicate the calculated percentages.

\subsubsection{Analysis for RQ2: Assessing Problematic Communication Practices about Described Youth Safety Features}
To examine potential misalignments in platform communication, two coders applied reflexive thematic analysis (RTA) \cite{Braun2019ReflectingAnalysis} inductively to analyze the quotation column in our dataset. RTA is a method for developing, analyzing, and interpreting patterns in qualitative data \cite{Braun2006UsingPsychology}, incorporating researchers’ pre-existing expertise and social positions. \textcolor{black}{We selected RTA over other methods, such as content or discourse analysis, for two primary reasons. First, our research goal was to identify and synthesize recurring themes of problematic practices across the dataset, instead of linguistic patterns or values in platform communication that discourse analysis can help surface \cite{Wodak2022CriticalAnalysis}. Second, while the communications are written for a general audience, uncovering problematic communication practices required a critical perspective that a general user's view alone might not provide. We therefore had to leverage our research team's expertise and backgrounds, making RTA the most appropriate choice as it formally incorporates this researcher positionality into the data interpretation, similar to how prior HCI work did so (e.g., \cite{Zhao2024OlderActivities, Ma2024LabelingYouTube}).} 

The first coder reviewed all entries in the quotation column and assigned initial codes in Google Sheets. For example, the code “Vague Risk Language in describing content risks” was assigned to a TikTok quotation: “Filtering out content with complex or mature themes from teen accounts, powered by our Content Levels system.” After this, the first coder consolidated similar codes into sub-themes and grouped related sub-themes into primary themes using a Miro board. In this process, the second coder regularly discussed all codes, associated quotations, sub-themes, and primary themes with the first coder to address disagreements and refine coding. This resulted in a thematic scheme (see codebook in Appendix~\ref{reflexive_codebook}), answering RQ2 from Sections \ref{section2.1} to \ref{section2.3}. 

\textbf{Researcher Positionality.} Our team's diverse positionality supports the critical researcher angle adopted in our analysis, as we acknowledge that youth online safety is an often politicized and emotionally charged topic. This angle leverages our collective academic expertise and varied perspectives from computer science, HCI, and social sciences as an international team of academics. Also, our perspectives are independent of the social media technology industry, as none of us is funded by these companies. Furthermore, some members of our team are parents, adding a personal dimension to our understanding and engagement with youth online safety. This diverse academic and personal grounding, combined with our team's collective expertise in social media, platform algorithms, and youth online safety, was continuously reflected upon during regular discussions to refine our thematic schemes. These varied standpoints informed our assessment of platform communication practices and underpinned our RTA approach.

%% file: 4.Findings.tex
\section{FINDINGS}
This section introduces the youth safety problem areas that we identified on social media platforms, and illustrates them with examples for safety features that platforms claim address risks that arise in these areas (RQ1). We then analyze how platforms \emph{communicate} about these safety features and highlight the often vague and noncommittal nature of the communications (RQ2).

\subsection{Problem Areas with Known Risks and Described Safety Features for Youth (RQ1)}
In their communications, social media platforms present a large number of safety features that they claim address risks that arise on the platforms. We thus introduce a taxonomy to organize these claims and guide our discussion. Existing youth safety frameworks, such as the 4Cs \cite{Livingstone2021TheChildren}, and our proposed extension, the 5Cs, are organized around online risks to youth. However, platform communications usually revolve around described safety features, not risks. Safety features often correspond to a ``problem area'' on social media platforms, which we define as the contexts where risks can emerge on the platforms, which, in turn, can be linked to more than one risk. Therefore, instead of organizing communications around risks, we organize them around \emph{problem areas}.

\subsubsection{Problem Areas on Social Media Platforms}
\mbox{}\
\label{section4.1.1}

\noindent 
We identified seven problem areas: Content Exposure, Interpersonal Communication, Platform Usage Patterns, Data Access, Monetization, Content Creation, and Platform Access (see Methods Section~\ref{sec:methods_RQ1} for the codebook). We also added Parental Supervision as an eighth, cross-cutting category that is not strictly a problem area. We begin by providing an overview of example safety features within each problem area that platforms claim to offer. \textcolor{black}{For this, we counted the number of articles talking about a safety feature across time, source, and platforms within each problem area. These counts are descriptive aggregates of our dataset and give a trend indication of platform communication on the safety features in general.} Note that the described safety features do not represent an assessment of whether these features are implemented, function as intended, or mitigate the risks claimed to address. 

\begin{figure}[ht]
    \centering
    \includegraphics[width=0.95\linewidth]{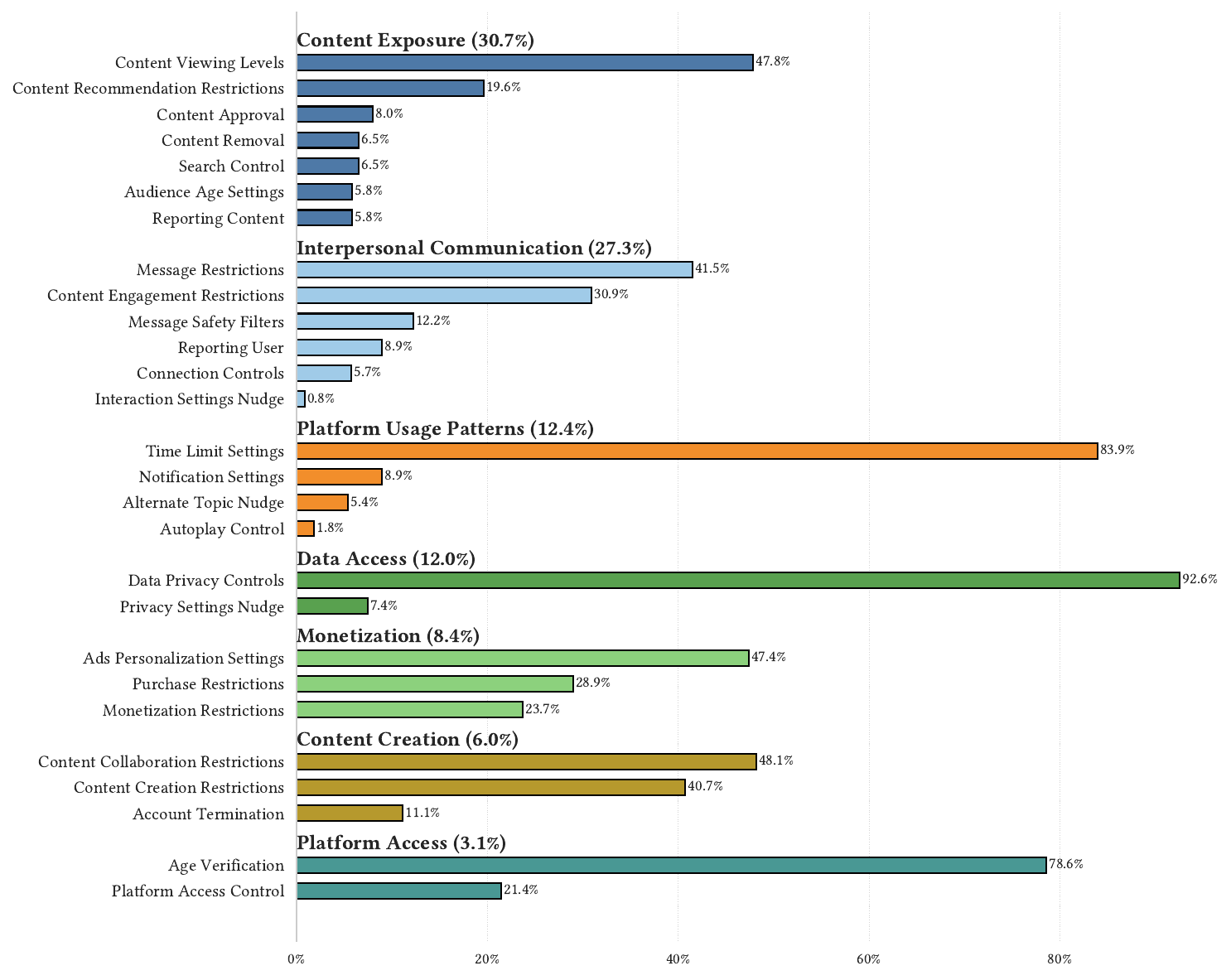}
    \caption{Problem areas comprising the described safety features with the frequency of their mentions.}
    \label{fig:barchart_problemareas_features}
\end{figure}

As shown in Figure~\ref{fig:barchart_problemareas_features}, \textbf{our findings show an uneven distribution of platform communications across problem areas}, where 30.7\% of communications focus on the Content Exposure area, while only 3.1\% focus on Platform Access. This highlights a strong emphasis of communication on moderating what youth \emph{see} rather than whether they can \emph{enter} platforms. Similarly, Interpersonal Communication (27.3\%) received substantial attention, while areas like Monetization (8.4\%) and Content Creation (6.0\%) were comparatively underrepresented. These patterns suggest that platform communication was concentrated around the interpersonal contexts, potentially overlooking the commercial or social contexts, where youth create and monetize content or perform commercial activities (e.g., purchasing products). Below, we detail each problem area.

\textbf{(1) Content Exposure:}
This is the most common problem area found in platform communications, which corresponds to the risk of youth encountering harmful, manipulative, or age-inappropriate content. The most prominent features mentioned within this area were \emph{Content Viewing Levels} (47.8\%) and \emph{Content Recommendation Restrictions} (19.6\%). 
These features reflect a strategy of top-down control, from either parents or platforms, in filtering what youth are allowed to see. 
For example, Instagram stated: ``Teen Accounts are automatically set to see less sensitive content in search results and recommended content in Explore, Reels and feed'' \cite{Instagram2024TipsParents}. This illustrates how platforms frame safety as something that can be managed through pre-set restrictions.

\textbf{(2) Interpersonal Communication:}
In this area, risks can emerge when communication channels such as messaging and commenting enable harassment, manipulation, or unwanted contact from peers or from adults. \emph{Message Restrictions} (41.5\%) were the most mentioned feature in the platform communications, followed by \emph{Content Engagement Restrictions} (30.9\%). The focus of these tools was primarily on \emph{restricting access} by controlling who can message whom or who can comment or tag. Notably, while platforms mentioned the introduction of nudging features, such as Instagram's comment warnings that prompt users to reconsider posting offensive language \cite{Instagram2020Safer2020}, these tools remained reactive in nature.

\textbf{(3) Platform Usage Patterns:}
Platforms encourage continuous user engagement, which can lead to excessive or compulsive use. The most communicated feature in this area is \emph{Time Limit Settings} (83.9 \%). For instance, TikTok and Instagram mentioned automatic daily screen time limits of 60 minutes for teens \cite{Instagram2024TipsParents, Keenan2023NewTikTok}, prompting them to close the app or enter a passcode to request additional time once the limit is reached. Instagram also promoted features, such as ``Quiet Mode'' \cite{Instagram2023GivingInstagram}, which pauses notifications and summarizes what the user missed upon return. This feature aims to prevent excessive use through timers, reminders, or screen-time limits, as a nudging design.

\textbf{(4) Data Access:}
This area describes how people manage access to their personal data by others or the platforms. Nearly all communication in this area centered on the feature Data \emph{Privacy Controls} (92.6\%), which allows users to manage profile visibility, discoverability, or data sharing. Data access becomes problematic when the personal data of youth is accessible to others and can be used for stalking, identity theft, targeted harassment, or grooming. For example, youth accounts were typically claimed to be set to private by default by social media platforms, which limited who could see their content or discover them.

\textbf{(5) Monetization:}
Risks emerge in this area when youth are encouraged to spend money without understanding its value, or when youth creators generate income through their content, as this can lead to labor exploitation or performance pressure. Additionally, platforms often personalize advertising based on user behaviors, which can expose youth to age-inappropriate or persuasive ads, and youth may not be equipped to fully understand or critically evaluate them. The most frequently mentioned safety feature was \emph{Ads Personalization Settings} (47.4\%), followed by \emph{Purchase Restrictions} (28.9\%), and \emph{Monetization Restrictions} (23.7\%). These features are largely regulatory in nature (i.e., mandated by law) and aim to protect children from financial risks by reducing exposure to targeted advertising or limiting the possibility of spending money on in-app purchases. 

\textbf{(6) Content Creation:}
Social media platforms can pose risks when youth share content without fully understanding its visibility or permanence. Such behavior may result in unwanted attention, grooming, bullying, or reputational harm. Additionally, the desire for visibility and engagement can foster performance pressure and attention-seeking behaviors, such as posting risky or revealing content. \emph{Content Collaboration Restrictions} accounted for nearly half (48.2\%) of the communicated safety features in this area, followed by \emph{Content Creation Restrictions} (40.7\%). These features were designed to prevent others from remixing, downloading, or circulating youth content. The underlying logic is focused on mitigating content misuse after it is published. For example, TikTok described a policy that disables the ability to duet or stitch videos by users under 16, and sets the default for users aged 16–17 to ``Friends only'' with downloading disabled \cite{TikTok2021StrengtheningTikTok}. 

\textbf{(7) Platform Access:}
Gaining access to the platforms is the first step to experiencing social media. Risks in this area relate to the possibility that youth under the intended minimum age may access platforms without safeguards or in ways that bypass these safeguards. Platform Access had the lowest overall communication volume, but within it, \emph{Age Verification} was communicated the most (78.6\%). This feature aims to prevent underage users from entering the platform, mostly through self-reported age checks. For example, Instagram reported requiring users to input their birth date and may prompt them to verify their age if AI-based systems detect inconsistencies \cite{Instagram2021AskingBirthdays}. 

For completeness, we also include safety features that enable parents to mitigate risks in different problem areas. We categorized them under a cross-cutting layer we call \textbf{Parental Supervision}. The most communicated feature was \emph{Parental Monitoring} (61.8\%), followed by \emph{Parental Control Settings} (27.9\%), and \emph{Individual Kid Profile} (10.29\%). These features enable parents to observe or adjust their child’s platform use and privacy settings. The trend here is external oversight, where parents, rather than youth, are positioned as central agents of safety. 

Platform safety communication may reflect underlying safety feature design priorities, for instance, a greater focus on the Content Exposure area (30.7\% of mentioned safety features) and on the Interpersonal Communication area (23.8\% of mentioned safety features). In the next section, we examine how known risks emerge across these problem areas and how social media platforms communicate about using safety features to address those risks.

\subsubsection{Mapping Social Media Problem Areas to Known Risks Faced by Youth}
\mbox{}\
\label{section4.1.2}

\begin{figure}[ht]
    \centering
    \includegraphics[width=0.85\linewidth]{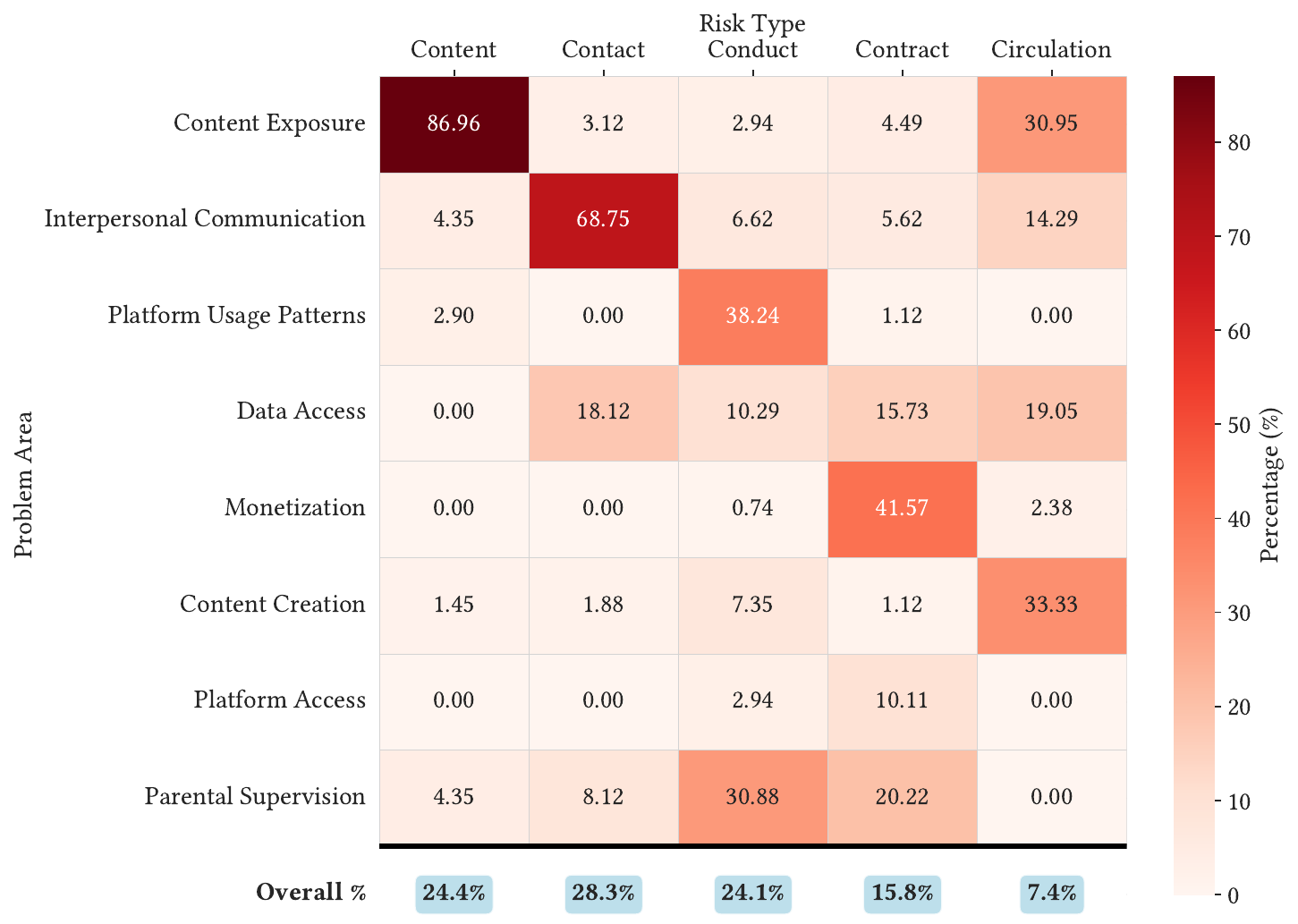}
    \caption{Distribution of the communicated safety features for each risk type across the problem areas. Problem areas are sorted from most communicated to least communicated. The percentage shows the relative frequency of the safety features across the problem areas that aim to address a risk. Each column sums up to 100\%. Overall percentages show how frequently risk types are mentioned by platforms.}
    \label{fig:heatmap_problemareas_risks_features}
\end{figure}

\noindent Using our 5Cs framework, this section explores how five risks, namely content, contact, conduct, contract, and circulation, emerge across the problem areas given by how platforms claim to address them (see Figure~\ref{fig:heatmap_problemareas_risks_features}). Please note that the heatmap does not represent the prevalence or severity of risks, but how frequently platforms allocate their communications across the problem areas, i.e., the co-occurrence of known risks and described safety features.

\textbf{Content risk} most arose in the Content Exposure area, as most safety features were communicated there. This suggests that platforms primarily frame content risk as a matter of what youth are exposed to, particularly through feeds, recommendations, or searches. Other areas where content may play a harmful role, such as Content Creation or Interpersonal Communication, were rarely linked to content risk in platform communications.

\textbf{Contact risk} appeared most often in the Interpersonal Communication and Data Access areas. This indicated that platforms primarily addressed this risk when it came from direct messaging, commenting, or being discoverable by others. This framing thus emphasized one-on-one, private interactions and underrepresents the peer-based or public dimensions of contact risk that often occur through Content Exposure or Content Creation.

\textbf{Conduct risk} largely emerged in the Platform Usage Patterns area and claimed to be addressed through Parental Supervision. Here, platform communications centered on tools like Time Limit Settings and Parental Monitoring, which suggested a behavioral framing that connects youth safety to screen time and adult oversight.

\textbf{Contract risk} was frequently associated with Monetization, Data Access, and Platform Access areas. Platforms often cite features like Age Verification, Purchase Restrictions, and Ads Personalization Settings. 
These protections often relied on self-reported information, which implied that contract risks are mitigated through gatekeeping and usage policies. 

\textbf{Circulation risk} arose across multiple problem areas, most notably Content Creation and Content Exposure. Platform communications reference various features to address this risk. For example, in Content Exposure, YouTube stated that they aimed to reduce the recommendation of risky content from Youth: ``We expanded our efforts from earlier this year around limiting recommendations of borderline content to include videos featuring minors in risky situations. While the content itself does not violate our policies, we recognize the minors could be at risk of online or offline exploitation'' \cite{Instagram2024InstagramAccounts}. In Content Creation, Instagram, for example, communicated a feature that did not allow non-followers to remix the content of youth: ``Teen Accounts are automatically set to only allow tags, mentions, and content remixing by people they follow'' \cite{YouTube2019AnFamilies}.

Overall, we found that platforms mostly communicated about features that aim to address content risks (24.4\%), contact risks (28.3\%), and conduct risks (24.1\%). Specifically, features aiming to address circulation risks (7.4\%) were largely undercommunicated. These patterns offer context for RQ2, where we examine how platforms communicate about the implementation or effectiveness of their youth safety features.

\subsection{Problematic Communication Practices regarding Youth Safety Features (RQ2)} 
\begin{figure}
    \centering
    \includegraphics[width=1\linewidth]{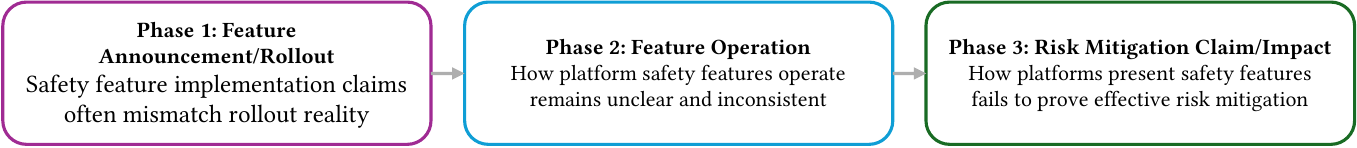}
    \caption{Problematic practices in social media platforms' communication lifecycle about youth safety features.}
    \label{fig:Communicationcycle}
\end{figure}

From an end-user standpoint, effective corporate communication should be clear, consistent, and easily actionable for a user to adopt their products \cite{Benedetto1999IdentifyingLaunch, Hultink1997IndustrialPerformance}. Thus, social media platforms ideally need to minimize users' cognitive load by clearly explaining risks and communicating about how features function to support their safety.
However, our analysis of platform communications related to youth safety found that the communications were vague and non-committal. We observed that such problematic communication practices followed a three-phase lifecycle, from (1) initial announcement to (2) describing the operational details of these features, and finally to (3) claims about their impacts. In the following, we describe these problematic practices across these three phases, where Figure~\ref{fig:Communicationcycle} provides an overview.

\subsubsection{Phase 1: Safety Feature Implementation Claims Often Mismatch Rollout Reality}
\mbox{}\
\label{section2.1}

\noindent This section addresses how platform communications about the implementation of safety features often mismatch the availability of those features. This misalignment arises primarily through two problematic communication practices. 

\textbf{(1)~Provisional and evolving language makes it unclear \textit{whether} features are implemented.} Platforms often announce features that they plan to implement or are testing, which can create uncertainty about whether or when these features are implemented. For example, in 2024, Meta repeatedly employed terms like ``\textit{testing},'' ``\textit{rolling out},'' and ``\textit{planning}'' to market its Nudity Protection feature in the Interpersonal Communication area (e.g., \cite{Meta2024CombatingNigeria, Meta2024IntroducingFacebook}). Such ambiguity can intensify when platforms announce feature testing without a concrete feature name, as when Meta described testing ``\textit{new pop-up messages for people who may have interacted with an account we’ve removed for sextortion…}'' \cite{Meta2024NewAbuse}.

\textbf{(2)~Stated geographic limits make it unclear \textit{where} features are implemented.}
Platforms often mention that they implement safety features only in certain regions, without stating a reason. For example, Snapchat described restricting personalized advertising in the Data Access area for teens, specifically ``\textit{in the EU and UK}'' \cite{Snapchat2023NewAct}, while TikTok described similar restrictions covering the ``\textit{European Economic Area, United Kingdom, and Switzerland}''~\cite{TikTok2023UpdatesTools}.
Neither mentioned such a restriction (or the lack of such a restriction) in the U.S. 
This geographically fragmented approach means that the described safety features might not be universally available, even after platforms promise a ``\textit{global rollout}.'' For example, Meta’s narratives about the Nudity Protection feature in the Interpersonal Communication area, starting from 2024, described it as ``\textit{rolling out globally}'' \cite{Meta2024NewAbuse}, while later noting it ``\textit{may not be available in your location right now}'' \cite{InstagramHelpCenter2025NudityInstagram}. Likewise, Meta stated in 2024 that its ``Teen Accounts'' were only active ``\textit{in some locations}'' with a full rollout planned in early 2025 \cite{Instagram2025ManageSettings}.
As a result, youth, parents (and researchers) must resort to testing for themselves whether a feature is available to them. This indicates an undue burden, and can be challenging for safety measures that operate in the background and have no permanently visible user interface components, such as the Nudity Protection feature.

\subsubsection{Phase 2: How Platforms' Safety Features Operate Remains Unclear and Inconsistent}
\mbox{}\
\label{section2.2}

\noindent This section discusses how problematic communication practices originate from a lack of clarity and consistency in describing how safety features operate. This manifests in three primary ways. 

\textbf{(1)~The conditions triggering safety features to activate remain unclear}, especially when conditions change without explicit notice and justification. The age threshold triggering safety features to work often lacks clarity, which can affect both children and parents. For example, Snapchat requires parents to be over 25 to utilize its Family Center \cite{Snapchat2023HowInformed} without explanation of why there is an age threshold for parents.
We observed that platforms usually do not mention a concrete rationale for age thresholds.
Furthermore, on several occasions, platforms have made changes to age thresholds without highlighting that the new thresholds represent a change or what motivated them to make the change. This can lead to confusion and inconsistency, particularly as the thresholds become stricter over time. For example, TikTok’s communications regarding default Private Account settings, in the Data Access area, evolved from applying to ages 13-15 \cite{Elizabeth2021OurTikTok} in 2021 to later stating, ``\textit{all accounts for teens under 18 start out as private}'' \cite{TikTok2025GuardiansGuide} in 2025. TikTok did not mention a rationale for the change, or even that there was a change. Following a similar pattern, Meta shifted from defaulting those ``16 and under'' in 2023 \cite{Meta2023HowExperiences} to ``\textit{under 18}'' in 2024 \cite{Instagram2024TipsParents} without full disclosure that this represented a change, and why the change was made. Furthermore, when age thresholds are combined with other factors, like whether users follow each other, it can become difficult for youth and parents to understand what specific conditions activate safety features.

\textbf{(2)~\textit{How} safety features intersect with each other remains unclear.} Platforms mention various safety features are available, but they often fail to clarify how the features intersect with each other or function together. 
For example, understanding the effectiveness of content filters is complicated when platforms mention multiple features without clear differentiation. TikTok mentioned ``\textit{Restricted Mode}'' \cite{TikTok2019TikToksParents}, a ``\textit{Content Levels system}'' filtering ``\textit{mature or complex themes}'' \cite{TikTok2022StrengtheningContent}, parental keyword filtering layered ``\textit{on top of our Content Levels system}'' \cite{TikTok2023UpdatingCouncil}, and separate ``\textit{maturity score[s]}'' \cite{Keenan2022MoreLove}. These all can create confusion about how safety features function together to mitigate content risks, which might eventually hinder users from adopting suitable features.

\textbf{(3)~Rationales for varying restriction levels regarding similar features remain unclear.} Platforms sometimes apply different levels of restrictiveness to features that may appear similar to users, but do not communicate why. For example, on TikTok, Duets and live streams are both forms of content creation, yet TikTok described different restriction levels: ``\textit{Only people over 18}'' can host a livestream \cite{TikTok2024TikTokCrisis}, but for Duets involving pre-recorded content, it highlighted that ``\textit{your teen has the control to decide who can make duets with or react to their content}'' to imply it was teens between 13 and 18 \cite{TikTok2019TikToksParents}.
It is likely that the platform assumes a different risk severity between live and pre-recorded content, but the communication failed to articulate this rationale well.

\subsubsection{Phase 3: How Platforms Present Safety Features Fails to Exhibit Effective Risk Mitigation}
\mbox{}\
\label{section2.3}

\noindent We found that platforms' presentations of safety features fail to demonstrate the effectiveness of claimed risk mitigation through four communication practices. 

\textbf{(1)~How platforms communicate about risks leaves the targeted risks unclear.} We found that platforms often fail to clearly explain the risks that safety features aim to address. For example, platforms often do not specify what they mean by inappropriate content. Meta described its Sensitive Content Control feature in the Content Exposure area as addressing potential ``\textit{harmful or sensitive content}'' \cite{Instagram2022HowInstagram} that ``\textit{may not be age-appropriate}'' \cite{Meta2023HowExperiences}, without offering a concrete characterization of the content targeted by these filters. Similarly, for contact risks, explanations can be circular rather than criteria-based: Meta defined ``\textit{potentially suspicious behavior}'' by adults not by describing the behavior itself, but by citing examples such as ``\textit{adults who have recently been blocked or reported by a young person}'' \cite{Instagram2024TipsParents}. Such vague descriptions also appear for contract risks, as when YouTube mentioned it ``\textit{won’t serve personalized ads or ads in certain categories}'' \cite{Beser2021AYouTube} in the Monetization area, without specifying which ad categories are not suitable for children.

\textbf{(2)~The effectiveness of safety features in addressing risks lacks evidence.} Platform communications about safety features' effectiveness often lack verifiable evidence. On the one hand, platforms explicitly admit that their safety features have limitations, such as content filters that can produce false negatives, stating that ``\textit{no system is perfect and inappropriate videos can slip through}'' \cite{YouTubeKids2019ResourcesParents}. On the other hand, when platforms do claim improvements in their safety features, the communication may lack verifiable evidence: YouTube asserted a new content classifier ``\textit{will detect and remove 2X more individual [predatory] comments}'' \cite{YouTube2019MoreYouTube} without the necessary baseline or context to validate this numerical claim. This lack of evidence-based reporting made it impossible to hold platforms accountable for their claims or to determine whether removing ``\textit{2X more}'' represents meaningful progress or an insignificant change.

\textbf{(3)~How platforms claim safety features work mismatches the scope of the risks they claim to address.} Platforms sometimes describe safety features in ways that mismatch the scope of the risks they purportedly address. For example, YouTube initially framed the ``\textit{Made for Kids}'' feature through the lens of policy compliance with COPPA (Children's Online Privacy Protection Act) \cite{YouTube2020FrequentlyKids} to address contract risks (i.e., protecting children from data collection without parental consent). However, its description stating creators are ``\textit{required to tell us that your videos are made for kids if you make kids’ content}'' \cite{YouTube2020FrequentlyKids} implied a broader application that might also address content and circulation risks, which the platform did not explicitly detail, and creates ambiguity for creators. Similarly, TikTok mentioned restricting the exchange of images or videos in Direct Messages \cite{Han2020ProtectingContent} but omitted mentioning other forms of content, such as ``\textit{effects, hashtags, or sounds [that can be sent] through a direct message}''~\cite{TikTok2025DirectMessages}.
This leaves it unclear whether these other forms of content are also restricted, or there is a possibility of circulation and content risks if they are exempt from the restriction.

In addition, platforms often avoid talking about risks by highlighting the opportunities their safety features offer. For example, Meta positively framed the safety feature Sensitive Content Control as a way for users to ``\textit{personalize their experience}'' \cite{Instagram2022UpdatesControl} and the Alternate Topic Nudge as designed to ``\textit{encourage teens to discover something new}'' \cite{Instagram2024AboutSettings}. Thus, this framing around personalization and discovery can misrepresent the full spectrum of risks platforms attempt to manage.

%% file: 5.Discussion.tex
\section{DISCUSSION}
Our study introduces the framework of \textit{problem areas} and investigates how platforms communicate about youth online safety. We uncover an unevenness of such communication: the mention of some risks, such as content and contact risks, is highly concentrated in a few problem areas, like Content Exposure and Interpersonal Communication, while other risks, especially conduct, contract, and circulation, are mentioned more sporadically across the problem areas (RQ1). In addition to such unevenness, we identified recurring problematic practices in platforms' communication lifecycle, spanning from feature/product rollout to operation and effectiveness (RQ2). In this sense, our study highlights the need to analyze platform communication practices, not just the safety features, as an integral part of youth online safety efforts by social media platforms. 

\subsection{Mapping Online Risks, Problem Areas, and Safety Features to Uncover Communication Gaps}
\label{DiscussionSec5.1}
While the prior 4Cs risk framework \cite{Livingstone2021TheChildren} provides a strong foundation for categorizing online risks faced by children (e.g., \cite{Badillo-Urquiola2024TowardsCare, Agha2023StrikePrevention, Wang2021ProtectionSafety}), understanding youth safety on social media platforms requires analyzing \emph{where} these risks arise and \emph{how} risks are claimed to be addressed by the platforms. Our problem areas framework contributes this missing layer by offering a structure for cross-risk and cross-platform analysis. This extends previous work that primarily focuses on the feature- or risk-specific levels (e.g., \cite{Agha2023StrikePrevention, Razi2020LetsExperiences, Ma2024LabelingYouTube, Dumaru2024ItsChildren}). Applying our framework to platform communications reveals a gap in how platforms communicate about the problem areas and describe safety measures.

Regarding \emph{content} risk, we found platform communication tended to focus on addressing the problem area Content Exposure over other areas. By narrowly focusing on passive content consumption, platforms overlook the risks of how youth create and share content. Also, platforms emphasize top-down content filtering by either parents or platforms. However, this contrasts with recent HCI and CSCW research showcasing youth's desire to shape their own content consumption experience \cite{Badillo-Urquiola2019StrangerOnline, Caddle2025BuildingSafety}.

In the context of \emph{contact} risk, we found that most communication prioritized restricting interpersonal communications with adults. While this aligns with traditional online safety narratives that focus on protecting children from online ``predators'' or strangers \cite{Livingstone2009BalancingSelf-efficacy, Shin2014ExploringSites}, this communicated focus underrepresents the contact risks that youth face in more public or peer contexts, such as Content Creation and Exposure areas. For example, peer-based harms like cyberbullying \cite{Chang2015TheDepression, Sasson2014ParentalAdolescents} can occur in public view within Content Exposure (e.g., via comments or public content creation), which extends beyond interpersonal interactions.

For \emph{conduct} risks, platform communication was concentrated on features like Time Limits and Parental Monitoring, which align with general recommendations on screen time for youth \cite{Legner2022KidsMuch} and prior work suggesting that awareness of children’s screen time can enhance parental mediation efforts \cite{RenkaiMa2025WeighingPlatforms}. However, risks that emerge through peer interactions 
such as peer aggression and cyberbullying \cite{Kirwil2009ParentalCountries, Chang2015TheDepression, Shin2014ExploringSites, Len-Rios2016EarlyLiteracy} are mentioned far less often. Through this imbalance, platforms signal a focus on individual self-regulation and parent-led oversight, mechanisms that are easy to measure and to promote. This neglects the social dynamics where mechanisms like peer support systems \cite{Razi2020LetsExperiences, Hartikainen2021SafeRisks} might help, but might be more complicated to communicate.

While platforms communicated about \emph{contract} risks across multiple problem areas, platform communication fails to address the challenges uncovered by recent HCI and CSCW work, which found that youth often lack awareness of datafication risks \cite{Goray2022YouthsMedia, Zhao2019IOnline, Sun2021TheyRisks} and that parents may be passive or underinformed about such risks \cite{Kumar2017NoPrivate}. As a result, platform communication related to the commercial and data-driven dimensions of contract risks can be described as superficial.

Lastly, we turn to \emph{circulation} risk, which we argue in this paper is not well represented in the existing 4Cs framework.
While youth increasingly act as content creators or influencers on social media platforms \cite{McRoberts2016DoYouTubers, Ruiz-Gomez2022PlayingInfluencers, Zulli2022ExtendingPlatform}, we found that platforms claim few features to address this risk, and these claims are scattered across different problem areas. As a result, users may not understand how these elements work together to mitigate the spread of sensitive or non-consensual content.

Illustrated across the five risks, we highlight the need to move beyond categorizing risks in isolation and toward analyzing the problem areas in which they arise. Our problem area framework supports contextual evaluation of platform safety communication by exposing how risk types intersect within specific areas of user activity on the platforms. By surfacing these cross-risk implications, our framework extends prior work that tends to focus on designing safety features against singular risks like contact \cite{Agha2023Co-DesigningInterventions, Agha2023StrikePrevention} or content \cite{Dumaru2024ItsTools}.

\textcolor{blue}{Beyond mapping these complex risks, the primary value of our problem areas framework is structural by providing a concrete taxonomy beneficial to future work in the auditing, comparison, and structured evaluation of safety features. For instance, future work can use our framework to conduct a direct audit of safety features within the Monetization problem area, comparing the claimed `Ads Personalization Settings' and `Purchase Restrictions' in Table \ref{tab:codebook_safety_features} across YouTube, TikTok, and Instagram. This provides a structure to identify gaps, as our findings for RQ1 revealed (e.g., the general neglect of the Monetization area). Furthermore, our problem area taxonomy can guide the design of experiments to test the effectiveness of safety features, such as comparing how well Instagram’s `Message Restrictions' in the Interpersonal Communication area performs against those of Snapchat. Ultimately, such context awareness, at the core of our framework, can facilitate recent efforts in HCI and CSCW toward evidence-based safety solutions for youth \cite{Obajemu2024TowardsNudges, Agha2023StrikePrevention, Badillo-Urquiola2024TowardsCare, Caddle2025BuildingSafety}, enabling contextual evaluation rather than reliance on platforms’ one-sided claims.}

\subsubsection{Communication Implications for Youth Safety Design}
\mbox{}\

\noindent \textcolor{blue}{\noindent While we did not directly study platforms' youth safety design philosophies, the way how they frame their communications hints at approaches that both echo and contrast with regulatory frameworks such as age-appropriate design codes \cite{InformationCommissionersOffice2021AgeServices} and evolving research like \textit{Safety by Design} principles \cite{Badillo-Urquiola2019StrangerOnline, Ghosh2020CircleFamilies}. The latter advocates for proactive and enabling solutions that integrate safety and risk prevention directly into the design and development of a product or services for youth. We briefly examine platforms' framing of youth safety across three dimensions: when interventions occur, who controls safety features, and how youth develop resilience:}

\textcolor{blue}{\textbf{(1) Proactive vs. reactive: When safety interventions take effect as communicated by the platforms.} Prior work has uncovered that youth online safety solutions are often reactive, intervening mostly after harm occurs \cite{Wisniewski2017ParentalSafety, Wang2021ProtectionSafety, Razi2021ADetection}. Our findings on platform communication present a nuanced picture. Platform communications were overwhelmingly dominated by proactive features, such as `Data Privacy Controls,' `Time Limit Settings,' and `Content Viewing Levels;' purely reactive features like `Reporting' were mentioned less (Figure \ref{fig:barchart_problemareas_features}). At the surface, an emphasis on proactive features appears to align with the principles of \textit{Safety by Design} \cite{Ghosh2020CircleFamilies}. However, prior research recommends that safety features should not only be proactive, but also be proactive in an enabling manner, as discussed below.}

\textcolor{blue}{\textbf{(2) Restrictive vs. enabling control: Who has agency over safety features as depicted by the platforms.} Our findings show that platforms have communicated more about safety features that centralize restrictive control of parents and platforms rather than giving agency to youth (e.g., Platform Access, Monetization, Platform Usage Patterns, Parental Supervision in Table \ref{tab:codebook_safety_features} and Figure \ref{fig:barchart_problemareas_features}).} This contrasts with recent research, which shows that children value personal privacy over parental controls \cite{Badillo-Urquiola2019StrangerOnline} and advocates for enabling mediation that guides technology use and keeps multiple stakeholders engaged \cite{Livingstone2017MaximizingMediation, Chen2016ActiveImpulsivity}. \textcolor{blue}{\textbf{There is an opportunity for platform communication to rebalance the safety narrative from one of control to one of guided autonomy.}
For example, in the context of account visibility in the Data Access area, platforms can provide age-appropriate explanations of the enabling options for youth, including benefits and risks of privacy settings.
In fact, however, we found that platforms obscured these youth-controlled settings through unexplained (and imposed, rather than enabling) shifts in the default settings per age group (Section \ref{section2.2}).} 

\textcolor{blue}{\textbf{(3) Passive vs. active learning: How youth develop resilience as communicated by the platforms.}} Current platform communication often positions youth as passive (e.g., subject to filtering or monitoring). Platforms used vague, undefined terms like ``sensitive content,'' which might prevent active learning (Section \ref{section2.3}). Furthermore, they positively framed these top-down controls as ``personalization'' rather than explaining the risks they mitigate (Section \ref{section2.3}). This passive framing in communication contrasts with the recognized shift in research towards strength-based approaches in online safety \cite{Ghosh2020CircleFamilies}, moving beyond protecting victims \cite{Badillo-Urquiola2021ConductingAdolescents} to empower youth with experience, knowledge, and skills in addressing risks \cite{ Zimmerman2013AdolescentPrevention}. \textcolor{blue}{\textbf{The implication is that platform communication itself should be an educational tool for building resilience.} For this, platforms should provide clear, age-appropriate explanations for why a feature exists and how it works. For example, an explanation for a Message Restriction feature should not only state that it blocks strangers, but also explain the types of contact risks (e.g., phishing, grooming) it helps to mitigate, thereby teaching youth risk assessment.}

\subsection{Platform Communication as Youth Online Safety Effort: Navigating Tensions in Enhancing Transparency}
\label{DiscussionSection2}
\begin{figure}
    \centering
    \includegraphics[width=0.7\linewidth]{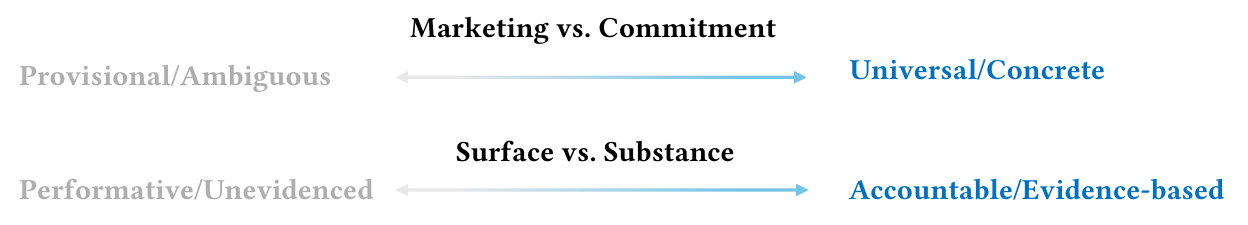}
    \caption{Two key tensions identified in platforms' communications for youth online safety.}
    \label{fig:Discussion2figure}
\end{figure}

Recent HCI and CSCW research has begun to evaluate the safety efforts of social media platforms, such as assessing the utility of current built-in features \cite{Dumaru2024ItsTools}, and co-designing new safety features with youth (e.g., \cite{Wang2023TreatOnline, Agha2023StrikePrevention, Badillo-Urquiola2021ConductingAdolescents}). Complementing this line of work, our study is among the first to evaluate how platforms communicate about safety features. How platforms announce, describe, and market their safety features is an integral component of their overall safety efforts for youth. Our findings reveal problematic practices across the communication lifecycle: Claims about feature implementation mismatch with rollout reality; explanations of feature operation are unclear; and descriptions fail to demonstrate the effectiveness of features in mitigating risks. These practices point to underlying tensions platforms have to navigate between demonstrating their responsibility for youth safety and managing increasing public concerns like media scrutiny \cite{Rosenblatt2024SenateSafety, Vanian2024MetasThrough} and legal liabilities \cite{FederalTradeCommission2023FTCData, FederalTradeCommission2019GoogleLaw, FederalTradeCommission2019Musical.lyInc.}. Examining these tensions is essential, as corporate communication can significantly affect user perceptions and adoption of products \cite{Guiltinan1999LaunchOutcomes, Lee2003TheInnovativeness}. Our study uncovers two primary tensions in platforms’ communication: (1)~marketing goals versus commitment, 
and (2)~surface versus substance, as shown in Figure \ref{fig:Discussion2figure}.

First, platforms face a tension between their marketing goals and their commitments to youth safety. Our findings show that platforms periodically announced new safety features, but their announcements over time concealed the features' actual availability, and evolving criteria made it unclear what triggers those features to activate and work (Section~\ref{section2.1}). While such communication practices aligned with corporate marketing goals like trust-building and adoption \cite{Lee2003TheInnovativeness}, they took place amid increasing public and regulatory scrutiny over harmful content circulation \cite{Abraham2022ApplyingChallenges, Ward2021UploadingChallenges} and its severe consequences on youth (e.g., suicidal incidents \cite{Rosenblatt2024SenateSafety, Vanian2024MetasThrough}), which created a need for clear, consistent information for users \cite{Benedetto1999IdentifyingLaunch, Hultink1997IndustrialPerformance}. Prior work has shown limited platform policies concerning harmful content on social media \cite{Boerman2020DisclosingRelationship, Smit2020TheBehaviors}. Our study adds that platform communication rarely clarified \textit{when}, \textit{where}, or \textit{how} the feature enforces policies.

Second, platforms often described safety features positively by emphasizing personalization or discovery (Section~\ref{section2.3}), but they admitted limitations only vaguely or omitted evidence validating the features’ effectiveness (e.g., YouTube comment classifier; Section~\ref{section2.3}). This communication approach, likely aiming to build user trust \cite{Lee2003TheInnovativeness}, mirrors risk-downplaying strategies observed in the tobacco and e-cigarette industries \cite{OConnor2017PerceptionsPurchase}. This reflects a tension between needing to claim effectiveness for public acceptance versus avoiding evidence-based evaluation \cite{Rim2016DimensionsCSR} or creating legal liability (e.g., \cite{FederalTradeCommission2019GoogleLaw, FederalTradeCommission2019Musical.lyInc.}). This lack of substantiation in platform communications echoes how prior HCI and CSCW found that safety features were perceived as ineffective for addressing youth's real concerns \cite{Agha2023StrikePrevention, Alluhidan2024TeenUse}. It suggests, therefore, that a long road remains to be taken toward safety features that are demonstrably effective.

These two tensions ultimately center on the transparency of platform communication artifacts/channels, such as the press releases and safety blogs we analyzed. Transparency would fundamentally require social media platforms, as \textit{sociotechnical} systems, to be accountable \cite{Fung2007FullTransparency} and to offer visibility into their complexity \cite{Ananny2018SeeingAccountability, Flyverbom2016DigitalVisibilities}. Prior HCI and CSCW research shows that explanations from online platforms can enhance transparency \cite{Ma2023TransparencyGames} and potentially increase user trust \cite{Kizilcec2016HowInterface}. In the context of youth safety, full operational transparency may not always be the solution. Challenges like low adoption of current safety features by parents \cite{Akter2022FromEquals} or limited awareness of those safety features among youth \cite{Obajemu2024TowardsNudges} may not be resolved through more openness alone. As prior work suggests, detailed disclosure of platform safeguard mechanisms could aid evasion by users \cite{Niverthi2022CharacterizingEvasion} or allow bad actors to exploit system knowledge for grooming or distribution of child sexual abuse material (CSAM) \cite{Kamar2025CutieLink, Pereira2024Metadata-BasedMaterial}. This highlights an important tension in platform communication practices: balancing the benefits of communication transparency for empowering users against the ramifications of transparency (e.g., feature misuse, circumvention of safeguards). How platforms demonstrate to the public how they manage this tension, therefore, serves as the key indicator separating genuine safety efforts from reputation management.

\subsubsection{Guidelines for Communicating about Safety Feature Implementation and Effectiveness}
\mbox{}\

\noindent Given the problematic practices in platform communications in our findings, we also make recommendations for improving the platforms’ communication practices around safety features.

\textbf{(1) We advocate for a mandate for transparency on feature availability}. To counter the ambiguous or geographically inconsistent rollout claims as reported by our findings (Section~\ref{section2.1}), regulators such as the U.S. FTC could require platforms to clearly and accurately report \textit{where}, \textit{when}, and \textit{to whom} specific safety features are operationally available. \textcolor{blue}{From a design perspective, this transparency must extend beyond press releases into more in-app mechanisms. For example, social media platforms could implement a personalized ``safety dashboard'' within the app that confirms for a user (and their parent, if linked) which safety features are actually active on their specific account, in their region, at that moment.}

\textbf{(2) Clear explanations of feature operation and responsibility are required.} Addressing the opacity that we found in feature descriptions (Section~\ref{section2.2}), including unclear triggers, regulations should mandate consistent explanations of how safety features actually function. Clear operational details are necessary for users to utilize safety features and understand the lines of responsibility of operating those features between themselves and the platforms. 

\textcolor{black}{\textbf{(3) Use the problem areas framework to contextually evaluate safety communication beyond features.}} To address the performative safety claims lacking verifiable evidence (Section~\ref{section2.3}), policy can require evidence-based substantiation through standardized metrics or independent auditing protocols. \textcolor{blue}{This could involve ``red team'' scenario testing, where independent auditors use test accounts to systematically compare platform promises against real-world feature effectiveness. Recent reports on platforms like Instagram, for example, have shown that many safety features are ineffective or easily circumvented when tested this way \cite{Bejar2025TeenMinors}}. Moreover, recognizing that communication is an integral part of a social media platform's safety efforts, regulators should evaluate beyond the features to the communication itself. Instead of assessing communications in the abstract, regulators (e.g., FTC, state Attorneys General) can use our problem areas framework to systematically probe how clearly and accurately platforms explain risks and features within each specific context. For example, they could demand consistent explanations for features within often-overlooked areas like Monetization or Platform Access, ensuring the narratives comply with legal expectations. This would allow regulators to better demand alignment between what platforms claim about their safety features and how those features actually perform.

\textcolor{blue}{\textbf{(4) Design communications to bridge stakeholder collaboration gaps.} The communication gaps we found have practical, negative consequences for collaboration. For instance, the vague effectiveness claims (Section \ref{section2.3}) might give parents a false sense of security, leading to conflict when a teen knows a feature is ineffective. Platforms should thus design and co-design honest communication materials for each stakeholder. This could include ``one-page guides for educators'' that frame the risks in under-communicated areas like Monetization or ``conversation starters for families'' that are based on the actual (not marketed) functionality of features. By providing a shared understanding of the problem areas and safety features, platforms can facilitate collaboration between youth, parents, educators, and policymakers.}

%% file: 6.LimitFuture.tex
\section{LIMITATIONS AND FUTURE WORK}
Our study has limitations that can also inform future work. First, we analyzed how platforms communicated about their safety efforts, but we did not independently examine whether the safety features actually work the way they are described. Future work can investigate how well platforms' actions align with their statements through empirical research. Second, we counted how often certain topics (e.g., features, risks) were mentioned in articles to gauge communication practices. While this reveals what platforms emphasized, it does not indicate whether users saw these messages or if they had any impact. Plus, these counts can change depending on how platforms publish information, such as using lots of small posts versus one large one. Examining if people actually saw and understood such youth safety-related communication would be a valuable next step. \textcolor{black}{Third, for step 4 in Section \ref{sec_datacollection}, we acknowledge that such a snowballing sampling approach may over-represent some safety features that platforms themselves emphasized through their internal sources. However, our research aim is to comprehensively capture \emph{platform-communicated} claims on safety features. Hence, we consider this data enrichment step as construct-consistent rather than a threat to our data validity, similar to a backward snowball sampling in systematic literature review \cite{Wohlin2014GuidelinesEngineering}. Moreover, the snowballed articles constitute only about one-fourth of the final dataset, which makes them supplementary but not dominant in our dataset.} \textcolor{blue}{Fourth, we only looked at press releases, help center entries, and safety blogs on the platforms' official websites. We did not examine other important channels through which companies communicate with their users and regulators, such as in-app user interface tutorials or help texts, safety-related advertising, videos, podcasts, and regulatory reports, including risk assessments for the Digital Services Act, which can be a valuable next step. Fifth, we focus only on problematic communication practices of social media platforms, which can introduce selection bias in data collection. However, this gave us the opportunity to identify gaps and areas for improvement.} Lastly, our study focuses on the U.S. context as the social media platforms in our study are all U.S. companies, so future work can understand how platforms communicate their safety features across regions. 

%% file: 7.Conclusion.tex
\section{CONCLUSION}
Social media platforms face increasing scrutiny over youth online safety, but currently, understanding how they communicate about their safety features in addressing online risks for youth remains limited. Our study is among the first to comprehensively examine how popular social media platforms, including YouTube, TikTok, Meta (Instagram and Facebook), and Snapchat, frame and present their built-in safety features for youth protection, leading to a taxonomy of safety features along problem areas that describe where risks arise and where platforms apply safety features to address the risks. By analyzing $N$=352 press releases and safety-related blogs from these platforms, we mapped the described safety features to seven problem areas. Our analysis revealed an uneven distribution of platform communication across these areas. For instance, communications are heavily focused on Content Exposure and Interpersonal Communication, while largely neglecting Content Creation, Data Access, and Platform Access areas. Furthermore, we identified three problematic practices of how platforms communicate about safety features, including the discrepancies concerning feature rollout and operation, as well as the claimed versus potential effectiveness of these features in mitigating known risks. We argue for holding social media platforms accountable for responsibly communicating how they protect youth from different online risks.